# Possible gravitoelectric dipole moment of neutrinos and atmospheric and solar neutrino anomalies

Anatoli S. Kuznetsov

*Institute of Physics, University of Tartu, Riia Street 142, 51014 Tartu, Estonia*

*E – mail: anatoli@fi.tartu.ee*

**Abstract**

We have carried out quantum-mechanical calculations of transition dynamics between the left- and right-handed helicity states $\nu_L \leftrightarrow \nu_R$ of the atmospheric muon Dirac neutrino $\nu_\mu$, which has a gravitoelectric dipole moment (GEDM) interacting with the gravitational field of the Earth. It is shown that the asymmetry of the zenith-angle distribution of this neutrino can be explained by an interference of the two mutually perpendicular linearly polarized plane neutrino states with different phase velocities (difference in the respective refractive indices $\delta n_0 \approx 0.7 \times 10^{-22}$) inside the Earth taking into consideration its gravitational field. The solar neutrino deficit can also be explained by the electron neutrino $\nu_e$ helicity flip transitions $\nu_L \leftrightarrow \nu_R$ due to the interaction between GEDM of $\nu_e$ and the gravitational field of the Sun. The characteristic mass $M_c \sim 10^7$ GeV for the gravitational dipole of the neutral and charged leptons have been estimated.

PACS number(s): 12.10.Kt, 13.15.+g, 14.60.St, 95.30.Sf

# 1. Introduction

The discovery of the asymmetry of the zenith-angle distribution of the atmospheric muon neutrino [1-5] and the registration of the total flux of the solar neutrinos through their charge-current (CC) and neutral-current (NC) interactions with detectors at the Sudbury Neutrino Observatory (SNO) [6] have been very important events in neutrino physics in recent years. Significance of these discoveries lies in their close connection with the nature of neutrino – its mass spectrum and possible dipole moments (see also Ref. [7]) – which requires going beyond the scope of the Standard Model (SM) of elementary particles, where neutrino is considered as a point particle with zero charge and rest mass, having spin 1/2 and only one helicity state (the left-handed for neutrino and the right-handed for antineutrino).

The lack of a fundamental theory describing the structure and the spectrum of masses of charged and neutral leptons has necessitated to employ phenomenological models, among which at present the most popular are the so-called oscillation models [8]. In these models it is supposed that the three known types of neutrino flavors (those of electron, muon and tau neutrinos) are a coherent superposition of physical fields with different masses and, therefore, quantum-mechanical oscillation transitions occur between the states of different neutrino flavors with the probability depending on the squared mass difference and the angle of mixing. Thus, the atmospheric neutrino experiments in Super-Kamiokande (SK) may be described by the model of two-flavor vacuum oscillations $\nu_\mu \leftrightarrow \nu_\tau$ with a nearly maximal mixing and the squared mass difference $\Delta m^2 \approx 3 \times 10^{-3}$ eV$^2$ [1]. Other vacuum oscillation models with three-neutrino [9] and even with four-neutrino flavor mixing [10] have also been proposed. It is supposed that when a neutrino is moving through matter, the forward coherent scattering by the background matter induces a refraction index that can change the dynamics of the neutrino flavor oscillations (Mikheyev–Smirnow–Wolfenstein (MSW) effect [11]).

There exist also some other models for explaining the neutrino anomalies, alternative to the flavor oscillation models. In particular, it has been demonstrated [12] that in the case of the Dirac neutrino, having magnetic moment and interacting with the magnetic field of the Sun, the helicity flip transition $\nu_L \leftrightarrow \nu_R$ occurs in the right-handed ("sterile") state which, as assumed, has an extremely weak coupling to neutrino detectors because of lack of not only the CC but even the NC interaction. In our opinion, the extension of the Standard Model with incorporating the right-handed neutrino state is quite logical from considerations of the symmetry with charged leptons and for reasons of prospective resolving the problem of neutrino anomalies (see also Ref. [13]).

The gravitational interaction, being a common factor for atmospheric and solar neutrinos, may be important for searching solution for both these anomalies. The influence of the gravitational interaction on the neutrino flavor oscillation has been considered within the framework of the general relativity, supposing a violation of the equivalence principle [14] or a violation of the Lorentz invariance principle [15]. Further it was suggested that the rotation–spin coupling may also lead to the neutrino helicity flip transitions $\nu_L \leftrightarrow \nu_R$. However, due to a very low angular velocity of the Earth and the Sun, this mechanism is of no consequence to the solution of atmospheric and solar neutrino problems [16].

This article proposes an alternative model for solving both neutrino anomalies in the case of the Dirac neutrinos. We consider the interaction of the neutrino gravitoelectric dipole moment (GEDM) (analog to the electric dipole moment in electrodynamics) with a weak gravitational field, analogously to the electromagnetic interaction (see, e.g., Ref. [17]) and we have carried out a calculation of quantum mechanical dynamics of the helicity flip transition $\nu_L \leftrightarrow \nu_R$. In this model, the asymmetry in the zenith-angle distribution of the atmospheric muon neutrino can be explained by an interference of the two mutually perpendicular linearly polarized plane neutrino states, having different refractive indices inside the Earth in the presence of its gravitational field. The solar neutrino deficit can also be explained by the electron neutrino helicity flip transition $\nu_L \leftrightarrow \nu_R$ due to the interaction of its GEDM with the gravitational field of the Sun.

## 2. The model

We consider the Earth within the Newton gravitational theory as a sphere of constant density ($\rho \approx 5.5$ g/cm$^3$). In the laboratory frame with $z$ axis parallel to the neutrino velocity vector $\boldsymbol{v}$, one can obtain for the non-zero components of the gravitational field, $g_x$ and $g_z$, the following relations inside the sphere

$$g_z(t) = \frac{GMvt}{R^3}, \tag{1.1}$$

$$g_x(t) = \frac{GMb}{R^3}, \tag{1.2}$$

and outside the sphere

$$g_z(t) = \frac{GMvt}{\left(b^2 + v^2 t^2\right)^{3/2}}, \tag{1.3}$$

$$g_x(t) = \frac{GMb}{\left(b^2 + v^2 t^2\right)^{3/2}}. \tag{1.4}$$

Here $G$ is the gravitational constant, $M$ and $R$ are the Earth's mass and radius, $b = R \sin\theta$ where $\theta$ is the zenith angle (see Fig. 1), $v$ is the velocity of the muon neutrino $\nu_\mu$ and $t$ is the time. The time moment $t = 0$ is chosen as the instant when the neutrino passes through the midpoint $C$ of its path inside the Earth and when the $g_z$ component changes its sign.

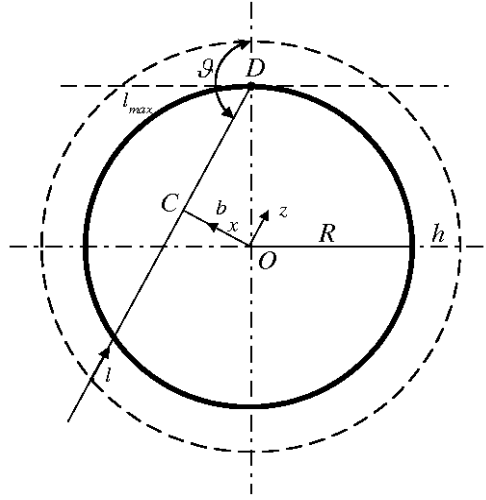

FIG. 1. Schematic picture illustrating the creation and registration of an atmospheric neutrino. Here $R$ is the Earth's radius, $h$ is an average height above the Earth's surface where the atmospheric neutrino is generated, $b = R \sin\theta$, $D$ is position of the neutrino detector, $l$ is the length of the neutrino trajectory in the Earth's atmosphere, and $\theta$ is the zenith angle.

Besides the neutrino path inside the Earth, we also take into account the neutrino path in the Earth's atmosphere after its birth at a height $h$ of about 15 km above the Earth's surface. From geometry it is easy to show that the length $l$ of a $\nu_\mu$ trajectory in the Earth's atmosphere is a symmetric function of the zenith angle relative to the horizontal direction ($\cos\theta = 0$), where it has the maximum $l_{max} \approx (2Rh)^{1/2} \approx 440$ km.

As is well known, the nonzero dipole moment of the elementary particle, collinear to the spin vector, may exist if both the parity ($P$) and the time reversal ($T$) invariance are simultaneously violated and as a result the $PT$ invariance is preserved [18]. Supposing that the neutrino has a non-zero GEDM, we consider only the case of the Dirac neutrino, because for the Majorana neutrino the existence of the dipole moment (diagonal with respect to the neutrino flavor) is impossible for the symmetry reasons. In addition, assuming that GEDM of the Dirac neutrino appears from a particle-antiparticle pair, we arrive at violation of the charge conjugation *C* and, consequently, also at a global *CPT* invariance violation (see also Ref. [19]).

The Hamiltonian describing an interaction between GEDM of the neutrino $\boldsymbol{d} = d\boldsymbol{\sigma}$ (here σ is the Pauli matrix) and a weak gravitational field $\boldsymbol{g}$ in the nonrelativistic limit can be written by analogy with electromagnetic interaction [20] in the form of the pseudoscalar product $H = -d\boldsymbol{\sigma}\boldsymbol{g}$. Then in the relativistic case the dynamics of the neutrino helicity flip transitions $\nu_L \leftrightarrow \nu_R$ under the influence of a weak gravitational field can be described by the Hamiltonian matrix (see also Refs. [12, 20, 21]):

$$i\hbar \frac{d}{dt}\begin{bmatrix} C_L \\ C_R \end{bmatrix} = \text{sign } g_z(t) \begin{bmatrix} 0 & idg_x(t) \\ -idg_x(t) & 0 \end{bmatrix}\begin{bmatrix} C_L \\ C_R \end{bmatrix} \qquad (2)$$

Here $C_L$ and $C_R$ are the amplitudes of the left-handed and right-handed states of neutrino and $\hbar$ is the Planck constant. In the Hamiltonian matrix we omitted the constant diagonal matrix elements with neutrino energy (which do not influence transition dynamics) and neglected second-order diagonal terms. The factor sign $g_z(t)$ takes into account the change in direction of the quantum transition (absorption ↔ emission of the gravitational radiation) between the energy levels of the neutrino left-handed and right-handed helicity states at the moment $t = 0$ when $g_z(t)$ changes its sign (by using an analogy with electromagnetic interaction in the magnetic resonance [22]). Such a (very small) energy difference $\Delta E \approx d\, g_x(t) \sim 10^{-11}$ eV is due to interaction of the neutrino's GEDM with the Earth's gravitational field. As we can see later, introduction of this factor, sign $g_z(t)$, provides a natural reversible helicity dynamics of neutrino relative to the center of symmetry of its trajectory inside the Earth.

To integrate the system given by Eq. (2) we must specify the initial values of the helicity states of $\nu_\mu$ at the moment of their birth in the Earth's atmosphere. We take them as $|C_L(t_i)|^2 = 1$ and $|C_R(t_i)|^2 = 0$ according to SM. Then the analytical solution of Eq. (2) is

$$|C_L(t)|^2 = \cos^2 \varphi(t), \qquad (3.1)$$

$$|C_R(t)|^2 = \sin^2 \varphi(t), \qquad (3.2)$$

where

$$\varphi(t) = \frac{d}{\hbar} \int [\operatorname{sign} g_z(t)] \; g_x(t) \, dt. \qquad (3.3)$$

This solution has three essential properties. First, the expression under integral is an odd function of time and its integral in symmetrical limits is zero. Thus, the traveling of the neutrino $\nu_\mu$ inside the Earth does not change its helicity states and, consequently, only the path in the Earth's atmosphere gives the non-zero contribution to the change of the neutrino helicity states. Second, the dynamics of quantum transitions $\nu_L \leftrightarrow \nu_R$ in the relativistic case is independent of the neutrino energy. Third, the mixing between the left-handed and right-handed neutrino states is maximal.

## 3. Calculation results and discussion

We have calculated the temporal dependence of probability for the left-hand state $|C_L(t)|^2$ of the atmospheric muon Dirac neutrino $\nu_\mu$ for various trajectories (determined by the value of zenith angle $\theta$) by performing a numerical integration of Eq. (2) with the initial conditions $|C_L|^2 = 1$ and $|C_R|^2 = 0$. The results are presented in Fig. 2.

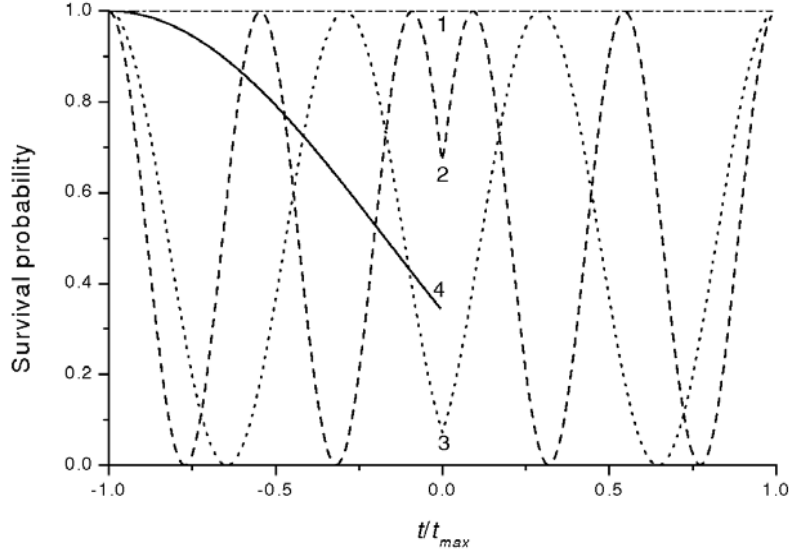

FIG. 2. The temporal dependence of the survival probability of the left-handed state for an atmospheric muon neutrino with GEDM value $d = 0.7 \times 10^{-27}$ g cm for different trajectories: *1* — $\cos \theta = -1$; *2* — $\cos \theta = -0.666$; *3* — $\cos \theta = -0.333$; *4* — $\cos \theta = 0$. The time scales for all curves are individually normalized to the time $t_{max}$ corresponding to neutrino birth in the Earth's atmosphere (at the initial point of the length *l* in Fig. 1).

For $|\cos \theta| = 1$ (the path of neutrino along the diameter of the Earth) the helicity state of $\nu_\mu$ does not change because $g_x(t) = 0$. In the interval $-1 < \cos \theta < 1$ helicity flip $\nu_L \leftrightarrow \nu_R$ transitions occur due to interaction of the neutrino's GEDM with the perpendicular component of the Earth's gravitational field. At the moment $t = 0$ when neutrino passes the midpoint of its trajectory inside the Earth, the time dependence of $|C_L(t)|^2$ has a critical point caused by the change in the sign of $g_z(t)$ in Eq. (2) due to the change in direction of the quantum transition. From considerations of symmetry of the Earth's gravitational field in relativistic case, there occurs a complete compensation of changes in helicity states of $\nu_\mu$ after its passage through the Earth. The exception is the portion of the $\nu_\mu$ trajectory in the Earth's atmosphere, which causes the symmetric (relative to the horizontal direction) zenith-angle dependence of the survival probability for the left-handed state of $\nu_\mu$ determined by the value of its GEDM (see Fig. 3).

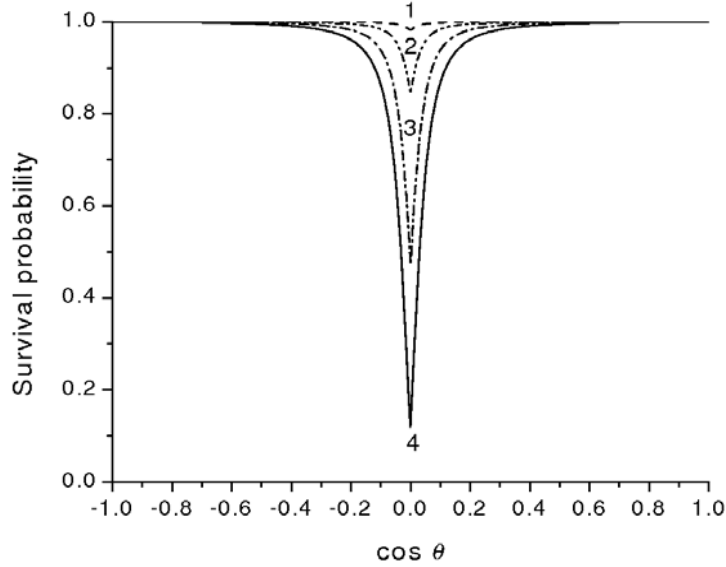

FIG. 3. Zenith-angle dependence of the survival probability for the left-handed neutrino interacting with the Earth's gravitational field and having different values of the gravitoelectric dipole moment (GEDM): $1 - d = 0.1 \times 10^{-27}$ g cm; $2 - d = 0.3 \times 10^{-27}$ g cm; $3 - d = 0.6 \times 10^{-27}$ g cm; $4 - d = 0.9 \times 10^{-27}$ g cm.

However, the experiments at SK [1] give an essentially asymmetric zenith-angle distribution of $\nu_\mu$ relative to the horizontal direction in a multi-GeV energy region. To explain this asymmetry in our model, we employ an analogy with the Pockels effect known in nonlinear optics. Due to the existence of only one perpendicular component of the Earth's gravitational field $g_x(t)$, the two mutually perpendicular linear polarization components of $\nu_\mu$ inside the Earth may have different refractive indices, so that $\delta n = n_x - n_y \neq 0$. Then the interference between those two linear polarization components of $\nu_\mu$, which is dependent on the neutrino energy, the length of neutrino trajectory inside the Earth and the intensity of the perpendicular component of the Earth's gravitational field, causes an oscillating dependence for the amplitudes of the left- and right-handed neutrino states.

The difference in the refractive indices as a function of a zenith angle, $\delta n(\cos\theta)$, for the two mutually perpendicular linear polarization components of $\nu_\mu$ in the presence of a perpendicular component of the gravitational field $g_x(t)$, averaged over the neutrino trajectory inside the Earth, can be approximated by

$$\overline{\delta n(\cos\vartheta)} = \delta n_0 \frac{\overline{g_x(\cos\vartheta)}}{g_x(\cos\vartheta=0)} = \delta n_0 \sin\vartheta . \tag{4}$$

Here $\delta n_0$ is a constant and $\overline{g_x(\cos\vartheta)}$ is the zenith-angle dependence of the perpendicular component of the Earth's gravitational field averaged over the neutrino trajectory inside the Earth. Note that in our case of a constant Earth's density the perpendicular component of the gravitational field, $g_x(\cos\theta) \propto \sin\theta$, remains constant all through the neutrino trajectory inside the Earth.

A neutrino, which is in its right-handed or left-handed helicity state and is traveling through the birefringent medium in the $z$ direction, can be represented as a coherent superposition of the two mutually perpendicular linearly polarized along $x$ and $y$ axes plane states

$$\left|\psi_{R,L}(z)\right\rangle = \frac{1}{\sqrt{2}}\left[\left|\varphi_x\right\rangle \exp(i \cdot k \cdot n_x \cdot z) \pm i\left|\varphi_y\right\rangle \exp(i \cdot k \cdot n_y \cdot z)\right]. \tag{5}$$

Then the survival probability of the left-handed neutrino helicity state inside the matter is given by

$$P_{LE}\cos\vartheta = \left|\left\langle\psi_L(0)|\psi_L(z)\right\rangle\right|^2 = 1 - \sin^2\left(\frac{\pi \cdot z(\cos\vartheta)}{L_{osc}}\right), \tag{6}$$

where

$$L_{osc} = \frac{2\pi}{k\delta n(\cos\theta)}, \qquad (7)$$

is the neutrino helicity oscillation length in the birefringent matter, $z(\cos\theta)$ is the length of the neutrino trajectory in the matter, and $k$ is the neutrino wave vector. It is important to note that dependence of the oscillation length $L_{osc}$ on the neutrino wave vector in our case is different from the case of the mass mixing, where $L_{osc} \propto k$. Taking into account the survival probability for the left-handed state of $\nu_\mu$ in the Earth's atmosphere, $P_{LA}(\cos\theta)$, we find the total survival probability for the left-handed state of $\nu_\mu$ in the form

$$P_L(\cos\vartheta) = P_{LA}(\cos\vartheta)\left[1 - \sin^2\left(\frac{E_\nu \overline{\delta n(\cos\vartheta)} R \cos\vartheta}{\hbar c}\right)\right], \qquad (8)$$

where $E_\nu$ is neutrino energy.

Supposing that $L_{osc} \approx 2R$ and $\delta n(\cos\theta) \approx \delta n_0$, we obtain from Eq.(7) that $\delta n_0 \approx 0.65\pi\hbar c/(E_\nu R)$. Using SK data [1], where it was shown that the asymmetry of the zenith-angle distribution of $\nu_\mu$ sharply increases with the rise of the neutrino energy at $E_\nu \geq 1$ GeV, we arrive at the estimation $\delta n_0 \approx 0.7\times 10^{-22}$.

Further, for the determination of GEDM of $\nu_\mu$ we used the data from the long baseline ($L = 250$ km) neutrino experiment KEK-to-Kamioka [23] (where 44 neutrino events *versus* 63.9 expected were observed). Supposing that in this experiment the sign of the $g_z(t)$ component of the gravitational field of the Earth did not change along the neutrino trajectory and using Eq. (3) we obtained for GEDM of $\nu_\mu$ the value $d = 0.7\times 10^{-27}$ g cm. Here we neglected the effect of birefringence for the muon neutrino by its passing through the Earth, because the helicity oscillation length for neutrino having an energy $E_\nu \approx 1$ GeV is very large ($L_{osc} \approx 6000$ km) as compared to the experimental baseline ($L = 250$ km).

Analogously, using the first results of the long-base (about 180 km) reactor experiments with an electron antineutrino in KamLAND [24], we estimated the value of its GEDM to be $d = 1.2 \times 10^{-27}$ g cm.

In addition, using Eq. (8) and obtained by us parameters ($d = 0.7 \times 10^{-27}$ g cm, see above, and $\delta n_0 = 0.65 \times 10^{-22}$, see below) for the atmospheric muon neutrino, we calculated the zenith-angle distribution of survival probability for the left-handed state in sub-GeV and multi-GeV energy ranges (see Fig. 4 (a) and (b)). The points represent the normalized (at $\cos\theta = 1$) data obtained in SK [1]. The calculations were averaged over the neutrino spectrum in the energy intervals 0.2÷1.4 GeV for sub-GeV and 1.4÷10 GeV for multi-GeV region and also averaged over the zenith angle with the step $\Delta(\cos\theta) = 0.2$. The calculations show an increase of the asymmetry in the zenith-angle distribution relative to the horizontal direction for the multi-GeV energy region of $\nu_\mu$ in agreement with SK data. Due to the difference in the refractive indices $\delta n (\cos\theta)$ for the two mutually perpendicular linearly polarized components of $\nu_\mu$ in the presence of the only one perpendicular component of the Earth's gravitational field, the normalized integral flux of $\nu_{\mu L}$ decreases from 1 to 0.86 for sub-GeV and from 1 to 0.71 for multi-GeV region, which can be compared with SK data 0.63 and 0.68, respectively [1]. The fitting of the SK data inside the sub-GeV energy region allows us to get for the parameter $\delta n_0$ a refined value $\delta n_0 = 0.65 \times 10^{-22}$ in accordance with the above estimation $\delta n_0 \approx 0.7 \times 10^{-22}$. Calculations for the multi-GeV region are also in agreement with the data of the zenith-angle distribution for the high-energy (about 10 GeV) upward-going muons produced by $\nu_\mu$ neutrinos inside the Earth [2,4]. Note that the averaging over the neutrino energy conserves the maximum at $\cos\theta = -1$ and the minimum at $\cos\theta = 0$ in the zenith-angle distribution of $\nu_\mu$. (see Fig. 4 (a)). The observation of these peculiarities in future atmospheric neutrino experiments with a high angle resolution would be a good test for the proposed model.

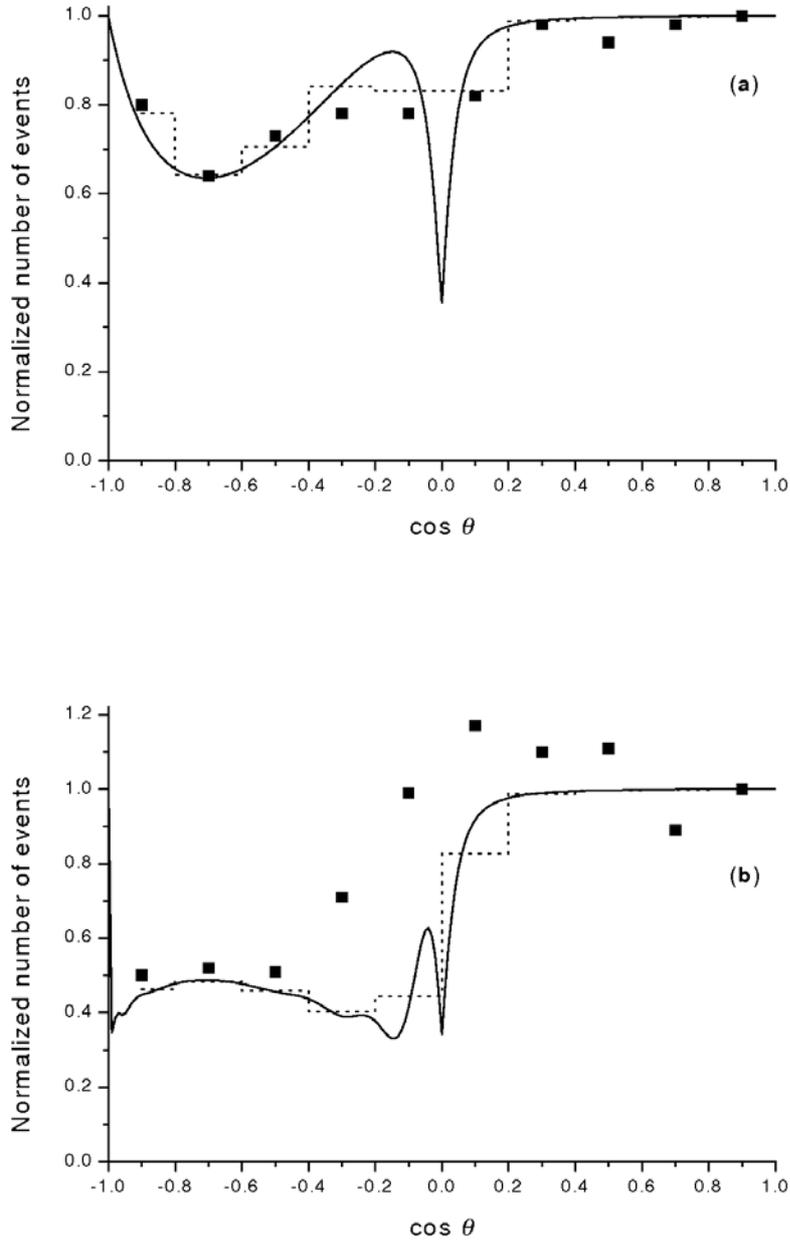

FIG. 4. Zenith-angle distribution of an atmospheric muon neutrino for the sub-GeV region (a) and the multi-GeV region (b) where the square dots represent the SK normalized data [1], the solid line is our calculated dependence and the dotted line is its histogram.

It is essential to note that the asymmetry of the zenith-angle distribution for atmospheric neutrinos $\nu_\mu$ relative to the horizontal direction in our model increases with growth of the neutrino energy because $L_{osc} \propto 1/E_\nu$ (see Eq. (8) and Fig. 4 (a) and (b)). This is in a good agreement with SK experiment [1] and, on the contrary, is in disagreement with prediction of the mass mixing oscillation model where $L_{osc} \propto E_\nu$.

The symmetric (relative to the horizontal direction) zenith-angle distribution of the atmospheric electron neutrino obtained in SK [1] allows us to estimate only the upper limits of the parameters $\delta n_0$ and $d$ for $\nu_e$, namely $\delta n_0 \leq 10^{-23}$ and $d \leq 10^{-27}$ g cm.

In our opinion, the solar neutrino anomaly can also be explained within the framework of the proposed model. Using Eq. (3) we calculated the dependence of the averaged survival probability of left-handed helicity state $\overline{|C_L|^2}$ for $^8$B electron neutrino on the magnitude of neutrino GEDM after spatial averaging over the central part of the Sun (where neutrinos are generated in the CC reactions). For the spatial averaging we used data of numerical calculations in the standard solar model (SSM) [25] to obtain the approximation function for the radial distribution of the Solar mass $m(r)$ and for the radial distribution of the $^8$B neutrino generation function $f(r)$:

$$\frac{m(r)}{M_0} = 1 - \frac{1}{a}\left(\frac{R_0}{r}\right)^3, \qquad (9)$$

and

$$f(r) = \exp\left(-\frac{r^2}{b}\right). \qquad (10)$$

Here $M_0$ and $R_0$ are the mass and radius of the Sun, while $a = 81.68$ and $b = 0.021 R_0^2$ are parameters. Calculations show (see Fig. 5) that increasing the magnitude of the $\nu_e$ GEDM causes decrease of $\overline{|C_L|^2}$ (due to helicity flip transitions in the gravitational field of the Sun).

For $d \approx 2.5 \times 10^{-32}$ g cm, the quantity $\overline{|C_L|^2}$ attains its minimum value of about 0.36 and further with increasing $d$, asymptotically tends to the expected value 0.5. Note also that this behavior of $\overline{|C_L|^2}$ in relativistic case is independent of the neutrino energy. The influence of the Earth's gravitational field on the helicity dynamics of solar $\nu_e$ can be neglected since it is much smaller than the effect of the gravitational field of the Sun (the phase changes in Eq. (3.3) for the $\nu_e$ GEDM value $d \approx 10^{-27}$ g cm are, respectively, $\varphi_E \approx 10^2$ and $\varphi_S \approx 10^4$). Therefore, in our model, the effect of the day/night asymmetry and also the season effect caused by ellipticity of the terrestrial orbit are both practically absent.

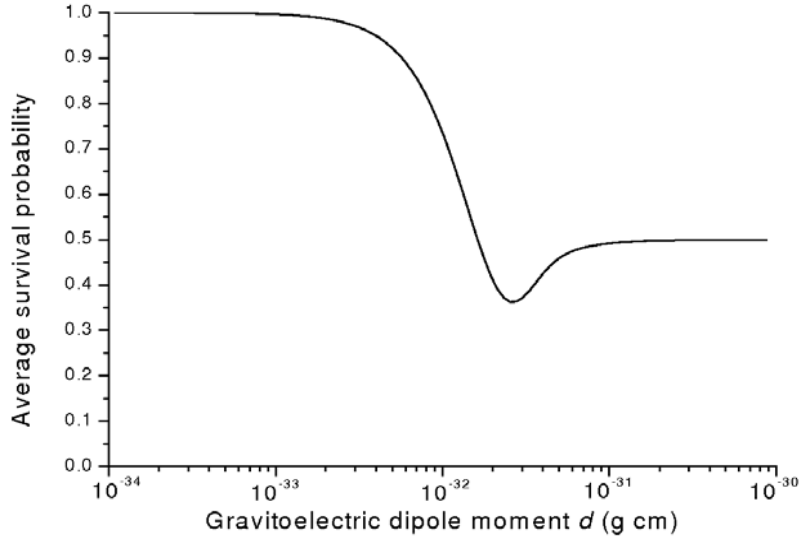

FIG. 5. Average survival probability of the left-handed helicity state of solar $^8$B electron neutrino as a function of the neutrino GEDM.

All these theoretical predictions are in good agreement with various experimental results in different solar neutrino energy regions. Thus, the SK data [1] of the ratio between the experimental value of the flux of the solar $^8$B neutrino and its theoretical value expected in the SSM for the elastic scattering reactions yields $R_{ES} = 0.47 \pm 0.08$. These experiments also revealed no distortion of the solar $^8$B neutrino spectra in the energy interval $5 < E_\nu < 20$ MeV and did not show the distinct day/night asymmetry or the season variation of the neutrino flux. Note also that in our model we have only one neutrino flavor $\nu_{eL}$ active in elastic scattering through CC and NC interactions with detector because in our case the muon and tau neutrinos are absent. Therefore, the ratio value for the ES reaction obtained in SK experiments is in good agreement with our predicted value 0.5.

Further, the solar electron neutrino radiochemical gallium experiments with low neutrino energy threshold ($E_\nu > 0.23$ MeV) for CC reaction of the $\nu_{eL}$ in the GALLEX [26] and SAGE [27] measurements give the ratio values $R_{CC} = 0.60 \pm 0.07$ and $R_{CC} = 0.52 \pm 0.07$, respectively, both being in close agreement with our prediction.

SNO Collaboration [6] detected the solar $^8$B neutrino through the CC, ES and NC reactions with the ratios $R_{CC} = 0.34 \pm 0.04$, $R_{ES} = 0.47 \pm 0.03$, and $R_{NC} = 1.0 \pm 0.13$ and did not find neither a distortion of the $^8$B neutrino energy distribution nor a distinct day/night asymmetry. The ratio value for the NC reactions can be explained in our model supposing (in addition to the SM) that both the left-handed and right-handed electron neutrinos may have equal NC weak interaction in the deuteron disintegration reaction $(\nu_e + d \rightarrow p + n + \nu_e)$. Note that the exotic scalar coupling of the right-handed neutrinos, in addition to the standard weak interaction of the left-handed neutrinos, was suggested in Ref. [28] as well. The ratio value for the ES reactions is in good agreement with our predicted value 0.5. The ratio value for the CC reactions can also be explained if the CC reaction data in the SNO experiments might be underrated by about 15% (for comparison, experimental data of neutrino-deuteron cross-sections have uncertainties of 10–30% [29]).

The first radiochemical measurements of the solar neutrino flux by the Nobel Prize Laureate (2002) Raymond Davis Jr. in the Homestake mine [30] give for CC reactions the ratio $R_{CC}$ = 0.34 which does not contradict our prediction if one takes into account statistical fluctuations of the Homestake data of about 20%.

Note that the possible explanation of the solar neutrino anomaly as a spin-flip transition $\nu_{eL} \leftrightarrow \overline{\nu}_{eR}$ (where $\overline{\nu}_{eR}$ is electron antineutrino) due to the gravitational interaction with the Sun in the case of Majorana neutrinos is excluded in view of the negative results of detecting the solar $\overline{\nu}_{eR}$ at KamLAND [31].

Interpretation of the solar neutrino anomaly as evidence of the neutrino flavor oscillation and MSW effect ($\Delta m^2 \approx 7 \times 10^{-5}$ eV$^2$ and nearly maximal flavor mixing [32]) requires a deformation of the $^8$B neutrino energy spectrum, an existence of the day/night asymmetry and emergence of muon and/or tau neutrinos; however, no one of these effects has been observed.

Based on the value of neutrino GEDM $d = M R \sim 10^{-27}$ g cm we can get estimation for the absolute value of two masses constituents of the neutrino gravitational dipole $M \sim d/R \sim d/\Lambda_c \sim 10^{-16}$ g $\sim 10^7$ GeV. Here we assumed $R \sim \Lambda_c = \hbar/m_e c \sim 10^{-11}$ cm ($\Lambda_c$ and $m_e$ are the Compton wavelength and the rest mass of the electron). The obtained mass value from the neutrino gravitational dipole is in a good agreement with estimation of the characteristic mass $M_c \sim 10^7$ GeV in the structure of the composite charged leptons obtained from the additional to the Standard Model contribution to the muon anomalous magnetic moment $\delta a \sim m_l/M_c$, where $m_l$ is the charged lepton mass [33]. To our mind, this remarkable agreement of the characteristic masses for the neutral and charged leptons is not accidental but, in opposite, it is naturally expected if proceed from the common structure of the whole lepton family.

Finally, our theoretical consideration of the difference in the refractive indices $\delta n_t$ for mutually perpendicular linear polarization components of the muon neutrino in presence of the gravitational field inside the Earth based on the gravitational polarizability of the rotating gravitational dipole of the electron in gravitational field led to the following expression

$$\delta n_t = \beta_e N_e \sim \frac{Gd^4}{\hbar^2 M} N_e, \tag{11}$$

where $\beta_e$ is the electron gravitational polarizability, $N_e$ is the average electrons concentration in the Earth. Using obtained parameters for $d \sim 10^{-27}$ g cm, $M \sim 10^{-16}$ g and suppose $N_e \sim 10^{24}$ cm$^{-3}$ we get estimation $\delta n_t \sim 10^{-21}$. It is in rather good agreement with the value $\delta n_0 \sim 0.7 \cdot 10^{-22}$ obtained above from the SK experimental data concerning on the asymmetry of the zenith-angle distribution of atmospheric muon neutrino, since $\beta_e \sim d^4$ uncertainty of $d$ strongly influences the estimated $\delta n_t$ value.

## 4. Conclusions

According to the suggested model for Dirac neutrinos, the atmospheric muon neutrinos and the solar electron neutrinos having gravitoelectrical dipole moments (GEDM) due to their interaction with the gravitational fields of the Earth and the Sun accomplish the helicity flip transitions $\nu_L \leftrightarrow \nu_R$. As a result of these transitions the flux of the left-handed solar electron neutrinos recorded by terrestrial detectors reduces twofold, independent of the neutrino energy. The asymmetry of the zenith-angle distribution of the atmospheric muon neutrino is explained by the interference of the two mutually perpendicular linearly polarized plane muon neutrino states having different refractive indices ($\delta n_0 = 0.65 \times 10^{-22}$) inside the Earth due to the perpendicular component of its gravitational field. The characteristic mass $M_c \sim 10^7$ GeV for the gravitational dipole of the neutral and charged leptons have been estimated. The gravitational interaction of Dirac neutrinos possessing GEDM violates $C$, $P$, $T$ and $CPT$ invariance. In addition to the Standard Model, we suppose that the right-and left-handed electron neutrinos may have the same NC interaction in the deuteron disintegration reaction.